\documentclass[twocolumn]{aa}
\usepackage{graphicx}
\usepackage{txfonts}
\begin{document}
\title{Chemical analysis of carbon stars in the Local Group:\thanks{Based on observations collected with the VLT/UT2 Kueyen telescope (Paranal Observatory, ESO, Chile) using the UVES instrument (program ID 71.D-0107)}}
\subtitle{I. The Small Magellanic Cloud and the Sagittarius dwarf spheroidal galaxy}

\author{P. de Laverny\inst{1}, C. Abia\inst{2}, I. Dom\'\i nguez\inst{2}, B. Plez\inst{3}, 
        O. Straniero\inst{4}, R. Wahlin\inst{5}, K. Eriksson\inst{5}, 
U.G. J{\o}rgensen\inst{6} }

\offprints{P. de Laverny; laverny@obs-nice.fr}

\institute{Observatoire de la C\^ote d'Azur, Dpt. Cassiop\'ee UMR6\,202, 06\,304 Nice Cedex 4, France
             \and
             Dpto. F\'\i sica Te\'orica y del Cosmos, Universidad de Granada, E-18\,071 Granada, Spain
             \and
              GRAAL, UMR5024, Universit\'e de Montpellier II, 34\,095 Montpellier cedex 5, France
              \and 
             INAF-Osservatorio di Collurania, 64\,100 Teramo, Italy
              \and
             Department of Astronomy and Space Physics, Box 515, 75120
             Uppsala, Sweden
              \and
             Niels Bohr Institute, Astronomical Observatory, Juliane Maries vej 30, 2\,100 Copenhagen, Denmark
}

\date{Received ; accepted }

 \abstract{We present  the first results of our  ongoing chemical study
of  carbon stars  in the  Local Group  of galaxies.   We  used spectra
obtained with UVES at the 8.2 m Kueyen-VLT telescope and a new grid of
spherical model  atmospheres for cool carbon-rich  stars which include
polyatomic  opacities, to  perform  a full  chemical  analysis of  one
carbon star,  BMB-B~30, in the  Small Magellanic Cloud (SMC)  and two,
IGI95-C1 and IGI95-C3, in  the Sagittarius Dwarf Spheroidal (Sgr dSph)
galaxy. Our  main goal  is to test  the dependence on  the stellar
metallicity  of the s-process  nucleosynthesis and  mixing mechanism
occurring  in AGB  stars. For these three stars, we find important s-element enhancements with 
respect to the mean metallicity ([M/H]), namely 
[s/M]$\approx$+1.0, similar to the figure found in galactic AGB stars
of similar metallicity. The abundance  ratios derived between  elements  
belonging  to  the  first  and  second  s-process abundance peaks, 
corresponding to nuclei
with a magic number of neutrons $N=50$ (88Sr, 89Y, 90Zr) and $N=82$
(138Ba, 139La, 140Ce, 141Pr), agree 
remarkably well with the theoretical predictions of low  mass 
(M $<3$~M$_\odot$) metal-poor AGB  nucleosynthesis models where the
main source of neutrons is the $^{13}$C$(\alpha,n)^{16}$O   reaction  activated during the long interpulse
phase, in a small pocket located within the He-rich intershell. The derived C/O and  $^{12}$C/$^{13}$C ratios 
are, however, more difficult to reconcile with theoretical expectations. 
Possible explanations, like the extrinsic origin of the composition of these
carbon stars or the operation of a non-standard mixing process during the 
AGB phase (such as the {\it cool bottom process}), are discussed on the basis
of the collected observational constraints.}

\authorrunning{P. de Laverny et {\em al.}}

\maketitle
%

\section{Introduction}

Asymptotic giant branch (AGB)  stars are  believed to be  the main producers  of 
s-elements in  the Universe. Indeed, the s-process is
activated  during the late  AGB phase,  within the  intershell region,
when  the  He-burning shell  suffers  recurrent thermal  instabilities
(thermal pulses or TPs).  After each TP, the convective envelope
can penetrate inward  in mass, dredging up the  material enriched with
the ashes  of He-burning (third dredge  up or TDU),  mainly carbon and
s-elements. Later on, this material is ejected through stellar winds,
modifying the chemical composition of the interstellar medium (cf. Iben
\& Renzini 1983; Busso et al. 1999).

In intermediate  mass AGB stars (M$\geq$3~M$_\odot$, IMS), 
free neutrons can be released at the base of 
the convective zone
generated by a thermal pulse through the $^{22}$Ne($\alpha,n$)$^{25}$Mg 
reaction (see e.g. Iben  \& Renzini  1983). Indeed,
in the He-rich intershell of these massive AGB,
the $^{22}$Ne is exposed to a temperature as high as $3.5\times 10^8$~K, 
which corresponds to
an equilibrium density of about 10$^{11}$ neutrons per cm$^{-3}$. 
However, due to such a high neutron density,
the resulting nucleosynthesis is characterized by the activation of 
several branchings
along the s-process path, leading to an elemental distribution of the 
heavy nuclei that is rather different from
those observed in the majority of the galactic MS, S and C (N type) stars
(see e.g. Lambert et al. 1995 and
Abia et al. 2000) and to an isotopic composition that is in conflict  
with those found
in meteoritic SiC grains, which are presolar condensates formed
in the outflows of carbon-rich AGB stars (e.g.  Zinner 1998).
On the contrary, in low mass AGB stars (M$<3$~M$_\odot$, LMS),
the temperature within the He-rich intershell barely attains $3\times 10^8$~K
and the $^{22}$Ne($\alpha,n$)$^{25}$Mg is only marginally activated.
Nowadays, it is well accepted that the $^{13}$C$(\alpha,n)^{16}$O reaction
is  the  main neutron  source  acting in  LMS.  
It only  requires  $\sim  10^{8}$ K  to  be activated,  a
temperature  usually  attained  at   the  top  layer  of  the  He-rich
intershell  during  the  interpulse  periods. The  extant  theoretical
models  (Straniero et  al.  1995;  Straniero et  al.  1997; Herwig  et
al. 1997;  Gallino et al. 1998;  Goriely \& Siess 2001)  assume that a
small amount of hydrogen is injected from the convective envelope into
the intershell region during the TDU.  At hydrogen reignition,
owing to proton  captures on $^{12}$C, a $^{13}$C  pocket forms within
the intershell. Then, the $^{13}$C is fully consumed by
$\alpha$ captures during the interpulse phase, and then leads to
a substantial s-process  nucleosynthesis with a
peak  neutron density  never  exceeding 10$^7$  cm$^{-3}$.  Then,  the
freshly synthesized s-elements  are engulfed by the convective shell
generated by  the next TP.   The marginal activation of  the $^{22}$Ne
neutron source  may slightly modify  the s-element distribution  within the
He-rich intershell.  The enhancement of the s-elements revealed by
spectroscopic studies of MS, S, SC  and C-stars confirms the occurrence of
repeated TDU episodes.

One  of the  most important  theoretical results  concerning  this new
s-process nucleosynthesis paradigm is its critical dependence on the
stellar metallicity and mass. Current models (e.g.  Busso et al. 1999;
Goriely \& Siess 2001)
show that  the predicted relative abundances of  the s-peak nuclei
(Zr, Ba  and Pb)  vary according to  the stellar metallicity.   At low
metallicity, the flow along the s-path  drains the Zr and Ba peaks and
builds  an excess  at the  doubly magic  $^{208}$Pb, which  is  at the
termination of  the s-path. Thus, as  the metallicity of  the AGB star
decreases, models predict larger Pb/Ba/Zr ratios. However, for a given
metallicity, a spread in the Pb/Ba/Zr ratios indicates a spread in the
amount  of  $^{13}$C, which  drives  the  neutron  production and  the
subsequent s-process nucleosynthesis in  a LMS (Gallino et al. 1998;
Delaude et al. 2004). The s-elements abundance pattern found in the 
metal-poor Pb-rich stars (Van Eck et al. 2001, 2003; Aoki et al. 2002)
and, at higher metallicities, in post-AGB stars (Reyniers
et al., 2004) 
points to the existence of such a spread.

A  comfortable  agreement  between  nucleosynthesis  calculations  and
s-element abundance patterns derived  in AGB stars of different types
in the  solar neighborhood  has been found  (e.g.  Busso et  al. 2001).  
However,  such comparisons mainly concern extrinsic stars 
(i.e.  stars that owe their  chemical peculiarities probably
to mass  transfer in a binary  system) belonging to  the disk of
our Galaxy.   Only in a few
cases has the predicted chemical pattern
in intrinsic AGB  or post-AGB stars (i.e. stars that owe their chemical peculiarities to 
an {\it in situ}  nucleosynthesis and mixing processes)
in a range of metallicity been  checked (Reyniers et al. 2004; 
Dom\'\i nguez et  al. 2004; Abia et al. 2002).   
Note that metal-poor
AGB belonging to the old galactic halo
have such small masses (0.6 $M_\odot$ on the average) that the TDU cannot 
take place (Straniero et al. 2003).

Due  to their  different chemical  evolution histories,  the satellite
galaxies have  stellar populations with metallicities  covering a wide
range ($-3.0\leq$[Fe/H]$\leq 0.0$, see e.g. Groenewegen 2004; Shetrone
et  al. 1998; Bonifacio  et al.  2004a). Therefore,  the study  of AGB
stars in the Local Group of galaxies provides an alternative sample of
intrinsic metal-poor  and relatively young AGB  stars. Most importantly,
since the distance of these external  stellar systems is well
known, a more accurate determination of  the stellar luminosity
can be estimated.  Furthermore, there is
observational  evidence  that  in  many  of these  galaxies  the  star
formation   histories  extended  at   least  until   a  few   Gyr  ago
(e.g. Demers, Battinelli \& Letarte 2003; Battinelli et al. 2003; 
Arimoto et al. 2004; Dom\'\i nguez et al. 2004; Rizzi et al. 2004). This  
increases the probability of observing in
these stellar  systems metal-poor  thermally pulsing AGB  (TP-AGB) stars, whose
abundance  pattern is  being modified  by  the occurrence  of the  TDUs
(intrinsic  AGBs).   Last  but  not   least,  the  knowledge   of  the
contribution  of  these low  metallicity  AGB  stars  to the  chemical
evolution  of their  parent  galaxies provides  new  hints to  discriminate
between   alternative  scenarios  of   galactic  formation   (Venn  et
al. 2004):  were the satellite  galaxies the basic building  blocks of
our own Galaxy? (e.g.  Bullock  et al.  2001), or are they debris of
larger  systems whose  structure and  evolution have  been  altered by
their proximity to our Galaxy (e.g. Grebel et al. 2003)?

In  this paper we report the first chemical  analysis of extragalactic
low-metalicity carbon stars found in the  Small Magellanic Cloud
and the Sagittarius dwarf spheroidal galaxy. 
The selected stars are  presented in Section 2 together with
the  observations. We  describe  in Section  3  the chemical  analysis
performed. Then, by analyzing the derived abundances and in particular
those of  the s-elements versus metallicity,  we provide constraints
to  the evolutionary  status  of  the studied  stars  and discuss  our
results in the framework of  the current AGB nucleosynthesis models at
low metallicity.

\section{Star selection and observations}

From the available catalogs  of carbon-rich stars in external galaxies
(e.g.  Groenewegen  \cite{groen}),  we first  selected  the
brightest   candidates  that   are  spectroscopically   confirmed  as
carbon-rich and as members  of their host galaxy.  Host galaxies
have been selected to span a large range of metallicities, in order to
provide better constraints to the nucleosynthesis models of AGB stars.
In  this first  study we  present the  spectroscopic results for three
carbon stars, one belonging to the Small Magellanic Cloud (SMC) and two
stars  to the  Sagittarius  dwarf spheroidal  galaxy  (Sgr dSph). 
Table~1 shows the photometric properties of our programme stars.

The SMC is an irregular galaxy at a distance of about 63~kpc (Cioni et
al.   \cite{cioni}) and suffering  an interstellar  extinction $A_{\rm
B}=0.16^{mag}$     (NASA/IPAC     database,     Schlegel    et     al.
\cite{schlegel}). Its mean metallicity is around [M/H]$=-0.74$
\footnote{Here we use the standard notation for the chemical abundance
ratio of any element X, [X/H] $=$ log (X/H)$_\star-$ log(X/H) $_\odot$
where  log (H)$\equiv  12$ is  the abundance  of hydrogen  by number. In this
scale the abundance of any element X is noted as $\epsilon$(X)}
(Luck et al. \cite{luck})  although lower metallicities (e.g. -1.05) are also
reported  (Rolleston  et al.  \cite{rolleston}).   The selected  star,
BMB-B~30 (B, standing for the bar of the SMC), was first identified as
carbon-rich by Blanco et  al. (\cite{blanco}).  Its carbon-rich nature
was later  confirmed by Rebeirot  et al. (\cite{rebeirot})  and, since
the survey  of Smith et  al. (\cite{smith}), it  was also known  to be
non-enhanced in lithium.

The  Sgr  dSph  galaxy  was  discovered  ten years  ago  by  Ibata  et
al. (\cite{ibata}).  It lies at about  26~kpc from the  Sun (Monaco et
al.   \cite{monaco})  and  is  characterized  by  a  large  spread  in
metallicity, $-0.8  \leq$ [Fe/H]  $ \leq 0.0$  (see e.g.  Bonifacio et
al. 2004b), and recent observations reveal the existence of stars with
metallicity  as   low   as   [Fe/H]  $\approx   -3$   (Bonifacio   et
al.   2004a).   The  adopted   interstellar   extinction  is   $A_{\rm
B}=0.66^{mag}$       (NASA/IPAC       database,      Schlegel       et
al.  \cite{schlegel}). The  two selected  stars were  confirmed  to be
carbon-rich  and  members   of  the  Sgr  dSph  galaxy   by  Ibata  et
al. (1995).  According to  Whitelock et al.  (1996), IGI95-C3, is 
one of the brightest star in Sgr dSph, with extremely red colors
suggesting a large mass-loss rate and/or
a low effective temperature (see below the problems encountered during
its chemical analysis). 

To date, the spectral type
of the three carbon stars is unknown since their full spectrum has
never been observed. We therefore cannot use common indicators as the relative intensity of
molecular bands of C$_2$ \& CN to classify them (such indicators lie
outside our observed spectral ranges). 
We have  estimated their  bolometric luminosity using the
calibrations by Alksnis et  al. (1998) from M$_{\rm{K}}$ (see Table~3).
These calibrations are obtained  from studies of galactic carbon stars
with known  distances from Hipparcos  parallax measurements. It  is not
clear whether these  calibrations can be safely applied  to metal-poor
carbon (C) stars  such as those  studied  here. Nevertheless,  using the  bolometric
corrections of M$_{\rm  K}$ from the (J-K)$_o$ index  for carbon stars
in the  SMC by  Wood et al. (1983),  we obtain  very similar
luminosities (see  Table~3). The bolometric magnitudes obtained
for BMB-B~30 and IGI95-C3  are compatible with the hypothesis that they
are TP-AGB stars undergoing TDU.  IGI95-C1 seems too  faint to be  an intrinsic 
C-star, even  considering that  the minimum luminosity  at which  an AGB
star  becomes carbon-rich is  lower at  low metallicity  (Straniero et
al. 2003). In fact, by using a simple core mass-luminosity relation to
constraint  the initial luminosity  of a  thermally pulsing  AGB phase
(Paczynski 1970),  and adopting M$_H=0.55$  M$_\odot$ for the  mass of
the H-exhausted  core, one gets a minimum bolometric  magnitude of
about $-4$. In any case,  the bolometric magnitudes shown in Table~3
are  affected by  large  uncertainties.  As pointed  out  by Busso  et
al. (2005),  with an effective temperature  of $\sim 3\,000$  K, most of
the  energy radiated by  a C-star  is in  the mid  infrared wavelength
range. By  using ISO  plus Two  Micron Sky Survey  data between  1 and
45~$\mu$m, these authors conclude that  the galactic carbon stars are, in many
cases, 1 or  2 bolometric magnitudes  brighter  than reported  in  
the available  catalogs (Groenewegen 2002).
 
 \begin{table*}
 \caption[]{Carbon stars observations log and properties.
References are: (a) Ibata et al. (\cite{ibata}), (b) Whitelock et al. 
(\cite{whitelock}), (c) Blanco et al. (\cite{blanco}) and
(d) Two Micron All Sky Survey.}
 \label{tab_log}
 \begin{tabular}[]{cccccccc}
   \hline
   \hline
   \noalign{\smallskip}
    Star &Date Obs. & Exp. time (min)& $B_J$ & $R$ & $J$ & $H$ & $K$ \\
   \noalign{\smallskip}
   \hline
   \noalign{\smallskip}
   Sag IGI95-C1 & 2003-07-03 & 195 & 16.9$^a$ &14.6$^a$ & 12.3$^b$ & 11.4$^b$ & 11.1$^b$ \\
   Sag IGI95-C3 & 2003-07-03/04/06& 195 & 17.3$^a$ &  14.3$^a$ & 11.3$^b$ & 10.0$^b$ &  9.4$^b$ \\
   SMC BMB-B~30 & 2003-06-11 \& 2003-07-05/30/31 & 195&  -- & 14.3$^c$ & 12.1$^d$ & 11.1$^d$ & 10.7$^d$ \\
   \noalign{\smallskip}
   \hline
 \end{tabular}
 \end{table*}

The three selected targets were observed in service mode with the UVES
spectrograph  attached to the  second VLT  unit (Kueyen  telescope) in
June and  July 2003.  The spectra  have been collected  using the UVES
standard  settings $Dichroic  2,  CD\#2$ centered  at  4\,370~{\AA}  and
$CD\#4$ centered  at 8\,600~{\AA}, leading to  observed spectral domains
from $\sim  4\,200$ to $\sim 5\,000$~{\AA} and from  $\sim 6\,700$ to
$\sim 9\,000$~{\AA}.  The  slit width was $1\arcsec$, corresponding to
a resolving power of about  40\,000.  The UVES Data Reduction Standard
Pipeline was  used for  the reduction of  the spectra. Then,  for each
star  and spectral  range, the  spectra were  first averaged  and then
binned by three  pixels as well as corrected to  the local standard of
rest. Finally, all the spectra were normalized to a local continuum by 
fitting a polynomial connecting the higher flux points in the spectral 
regions studied. For this procedure and, as a guide, we used the continuum location 
that may be inferred looking at the spectral atlas of carbon stars of different types
by Barnbaum et al. (1996) and, in particular, in the $8\,000$~{\AA} region 
the {\it theoretically} expected continuum windows identified by Wyller (1966) were used 
as reference points. Due to the huge number of atomic and molecular lines 
used for the synthetic spectra calculation, it is reasonable to think
that the theoretical continuum points are not too far from the
true continuum. Indeed, our theoretical spectra show maximum flux points
at these wavelengths. It was never necessary to modify the initial
placement of the continuum by more than $\pm 5\%$. Errors introduced by this
uncertainty in the continuum position were taken into account. However, systematic errors 
due to a larger uncertainty in the continuum location cannot be completely
discarded (see below).

\section{Analysis: The extragalactic carbon stars and their chemical properties}

We have  used the method  of spectral synthesis  in LTE to  derive the
chemical abundances  of the sample  stars with a special  emphasis on
specific spectral regions. In particular, the selected regions were: (i) 
between  $\,4750-4\,950$~{\AA}   mainly   for    s-elements   and    the   mean
metallicity\footnote{The mean metallicity  of the stars [M/H] (Table~3)
was obtained as  the mean value derived from several  lines of Fe, Ti,
V, Ni, Cr and Mn.}, (ii) $6\,700-6\,730$~{\AA} for Li, (iii) 
$7\,050-7\,080$~{\AA} for some Ti lines and one  Sr line, 
(iv) $7\,780-7\,820$~{\AA} for Rb and two
Ni lines and  $7\,990-8\,040$~{\AA} to derive  the carbon isotopic
ratio.  The adopted atomic line list is basically that used in Abia et
al. (2001, 2002). We have added  some lines taken from the atomic data
bases of Kurucz (CD-ROM No.  13) and VALD (Kupka et al. \cite{kupka}).
Some  revisions  have  been  made  using solar  gf-values  derived  by
Th\'evenin (1989,  1990), and the  theoretical estimates by Xu  et al.
(2003) and Den  Hartog et  al.  (2003)  for the Nd  II lines,  and the
DREAM data base (http://w3.umh.ac.be/~astro/dream.shtml) for Sm and Ce
singly ionized  lines. With  respect to Abia  et al. (2001,  2002), we
have identified additional spectral features corresponding to elements
whose abundances  are derived here (see Table~2).


\begin{table}
\begin{minipage}[t]{\columnwidth}
\caption{Spectroscopic parameters of the new identified lines.}             
\label{table:2} 
\renewcommand{\footnoterule}{}     
\centering          
\begin{tabular}{c c c c }     
\hline\hline       
                   
Ion & $\lambda$({\AA}) & $\chi$(eV) & log gf \\ 
\hline                     
Zn I& 4\,810.52& 4.08 & -0.170 \\
Nb I& 4\,802.44& 0.08 & -1.900\\
Ru I& 4\,757.84& 0.93 & -0.280\\
Ru I& 4\,769.30& 0.81 & -1.740\\
Ru I& 4\,844.56& 1.12 & -0.810\\
Pr II&4\,801.13& 0.48 & -0.878\\
Pr II&4\,837.03& 0.20 & -1.464\\
Pr II&4\,848.52& 0.05 & -1.584\\
Gd I &4\,758.70& 0.88 & 0.045\\
Hf I &4\,800.50& 0.70 &-1.210\\
W I& 4\,843.84 & 0.41 &-0.990\\
\hline                  
\end{tabular}
\end{minipage}
\end{table}


The molecular line list includes CN,  C$_2$, CH and MgH lines with the
corresponding   isotopic   variations. C$_2$ lines are from Querci et al.
(1971). CN and CH lists were assembled from the best available data and
are described in Hill et al. (2002) and Cayrel et al. (2004). MgH lines
come from Kurucz (CD-Rom No. 13). In particular, the MgH  lines, which were
not  included in  Abia et  al. (2001,  2002),  
act like a pseudo-continuum below  $4\,800$ {\AA}.   This might  be
significant   if   the   star   under   analysis  is   found   to   be
$\alpha$-enhanced ([Mg/Fe]$>0  $), as one would  expect for metal-poor
stars according to  the trend found in our Galaxy.  However, as far as
we know, all the chemical analysis so far performed in the SMC and the
Sgr    dSph   stellar    populations   do    not   reveal    a   clear
$\alpha$-enhancement trend. Thus, we have adopted in
the analysis no enhancement, i.e. [Mg/Fe]$=0.0$. 
The ratio [Ca/M] derived in two stars (see
Table~4) is compatible with this figure although we note that our Ca
abundance estimation is made only from one line, CaI 6\,717~{\AA}. 
We checked our line list  by comparing theoretical
with  observed  spectra  of  the  Sun and  Arcturus.  The
comparison  with Arcturus  allowed us  to check  some  molecular lines
whereas  most  of  these  features   are  not  visible  in  the  Solar
spectrum. For the Sun, we  used the semi-empirical model atmosphere by
Holweger \&  M\"uller (1974) with  abundances from Grevesse  \& Sauval
(1998).  For Arcturus  we computed a model  atmosphere with main
parameters according to Decin et al.  (2000) and
abundances  from Peterson  et  al.  (1993).  Theoretical spectra  were
computed  with the  turbospectrum code (Alvarez \& Plez 1998, and further improvements by  Plez) 
in spherical geometry.  The comparisons showed an
excellent agreement with the solar spectrum in all the spectral ranges
studied,  although the  comparison  was not  as  good in  the case  of
Arcturus in the $4\,750-4\,950$~{\AA} region. This indicates that
our  atomic and  molecular line  list is  still not  complete  in this
spectral range (despite  the fact that we included  about 100\,000 lines
in this region)  and/or the line data of  some molecules are partially
erroneous. Most of the spectral features used in the chemical analysis
of  our  stars,  nevertheless,  are  well  fitted  in  the  Arcturus
spectrum.

The   effective   temperatures  were   first   estimated  from   the
calibrations of the (V-K)$_o$  and (J-K)$_o$ color indexes by Aaronson
\&  Mould (1985). However,  the estimated  values are  just used  as a
starting  point,  the final  values  adopted  in  Table~3 being  obtained
through  an iterative  process by  comparing observed  and theoretical
spectra  computed with  different effective  temperatures. It  is well
known that AGB stars are variable and, hence, their T$_{\rm{eff}}$ may
change during  the pulsation. 
The photometric calibration
used  to   estimate  the  effective  temperature   indicates  that  the
uncertainty  in our T$_{\rm{eff}}$  is certainly  not lower  than 
$\pm 250$~K. We also note that effective temperature
estimates from (V-K) may be affected by stellar variability
since the measurements in the two bands were obtained on different occasions.
Nevertheless,  the final adopted
values  agree  quite well  with  those  derived  from the  photometric
calibrations within $\pm 100$~K. 
We set the gravity at log g$ = 0.0$ for all the stars, 
following Lambert et al. (1986) who to a considerable
degree based their choice on the luminosities and masses of 
carbon stars in the Magellanic Clouds. For the microturbulence,
we adopted $\xi=2.2$~km~s$^{-1}$, which is a value suitable for
AGB stars (Lambert et al.  1986).
Theoretical spectra were convolved using Gaussian functions with a
FWHM value according to the instrumental profile ($\sim 0.15$ {\AA}) and macroturbulence
velocities ranging from 4 to 7~k~s$^{-1}$. Uncertainties in the abundances
derived due to changes in gravity, microturbulence and macroturbulence within
$\pm 0.5$~dex, $\pm  0.20$~km~s$^{-1}$ and $\pm 1$~km~s$^{-1}$, respectively, 
were considered. 

Apart from T$_{\rm{eff}}$, the main atmospheric parameter affecting
theoretical spectra 
of carbon-rich stars is 
the C/O ratio. We estimated this ratio through an
iterative  process,  by  comparing  observed and  theoretical  spectra
computed for different values of the carbon abundance, while keeping
the other atmospheric parameters constant. For  that purpose,  we mainly
considered the region  near 8\,000~{\AA} since it is the most
sensitive to changes in the C/O ratio.  Indeed, the spectrum of C-stars  in this region is 
mostly affected  by CN and much weaker C$_2$ absorptions. 
We could not derive the carbon abundance directly from C$_2$ 
lines since (i) our linelist is not enough accurate around the 8\,000~{\AA} 
region (see Plez \& Cohen, 2005), 
(ii) the C$_2$ Swan (0,0) band around
5\,160~{\AA} was not observed and (iii) the collected spectra in the blue part
become too noisy
to consider the C$_2$ Swan (1,0) band around 4\,735~{\AA}.
We thus derive the carbon abundance from the CN lines
assuming that [N/Fe]$=0.0$. 
We checked in previous works
(Abia \& Isern 1996) that the C/O ratio derived from the 8\,000 {\AA} range is almost 
insensitive to the assumed abundance of nitrogen within $\pm 0.3$~dex.
However, the derived C/O ratios have an additional uncertainty due to the 
adopted O abundance which we cannot determine independently. This is because theoretical 
spectra are almost insensitive to a large variation in the O abundance provided that the 
difference $\epsilon$(C)$-\epsilon$(O) is kept constant. 
Indeed, when adding equal amounts of carbon and oxygen to the atmosphere,
the sole effect in the outer layers is to increase the abundance of 
the CO molecule
which has a negligible effect on the atmospheric structure.
Therefore, this ambiguity allows a 
range of oxygen abundances and C/O ratios giving almost identical synthetic spectra. 
Nevertheless, even considering an uncertainty of a factor three in the oxygen abundance, 
the C/O ratios derived in our stars are found in the range $1<$ C/O $<1.5$.     
On the other hand, and as  already  found  by  Abia  et
al.  (2002),  the  derived  C/O  ratio  slightly  decreases  with  the
wavelength of the spectral  region fitted.  For instance, in BMB-B~30,
we found  a best  value of C/O  $=1.2$ at  $\sim 8\,000$~{\AA},  but C/O
$=1.09$ at $\sim 4\,800$~{\AA}. Similar differences are obtained for the
other  two   program  stars.   The  reason  of   this  discrepancy  is
unknown.  It may be  related to  the continuous  opacity of  the model
atmosphere or to  an incomplete or erroneous molecular line list.
Note that a modification of the continuum location by $\sim 15\%$ in the
$4\,750-4\,950$~{\AA} range would increase our derived C/O ratio in this region 
by only a few hundredth of dex. Thus, the discrepancy would remain.  
The range in the C/O ratios derived for each star is given in Table~3 as well as the
corresponding abundance difference $\epsilon$(C)$-\epsilon$(O).

Regarding the model atmospheres, we used  a new grid of models for cool  carbon-rich stars.  
The spherically symmetric model atmospheres  were calculated with
the MARCS code, using opacity sampling in 11\,000 frequency points. Atomic,
diatomic and polyatomic (C$_2$H$_2$, HCN and C$_3$) absorptions were included.
The microturbulence parameter was set to 2-3 km/s. Turbulence pressures were
neglected. Convection was included according to the local mixing-length recipe
and found to be insignificant. The masses of the models were set to 
2~M$_{\odot}$, following Lambert et al. (1986). The abundance difference 
$(\epsilon(C)$-$\epsilon(O))$ and the overall metallicity were varied.
The adopted solar abundances for C, N and O in the models are 8.41, 7.80 and 8.67 
according to the recent revision by Asplund et al. (2005). As mentioned
previously, the models were found to be quite insensitive 
to $\epsilon(O)$ as long as $(\epsilon(C)$-$\epsilon(O))$ was unchanged. The typical 
extension (from optical depths 10$^{-4}$ to 10$^{2}$ in $\tau_{\rm{Ross}}$) 
was 7\% of the total stellar radius. The temperature structures for the models
used for the three stars are presented in Fig.~1. Further details will be 
given in forthcoming papers by Gustafsson et al.
(2005) and J{\o}rgensen et al. (2005). 
For each  star we chose from the grid a specific  
model atmosphere with a given value of T$_{\rm{eff}}$, C/O and the 
metallicity and  proceeded iteratively by
changing  these  parameters until  a  good fit  (by  eye)  in all  the
spectral regions  was found. Then,  the  abundances of  the
other chemical  species (s-elements, Li,  etc) were changed  to fit
specific spectral features.

\begin{figure}[ht]
  \begin{center}
\resizebox{\hsize}{!}{\includegraphics[angle=270,
width=\textwidth]{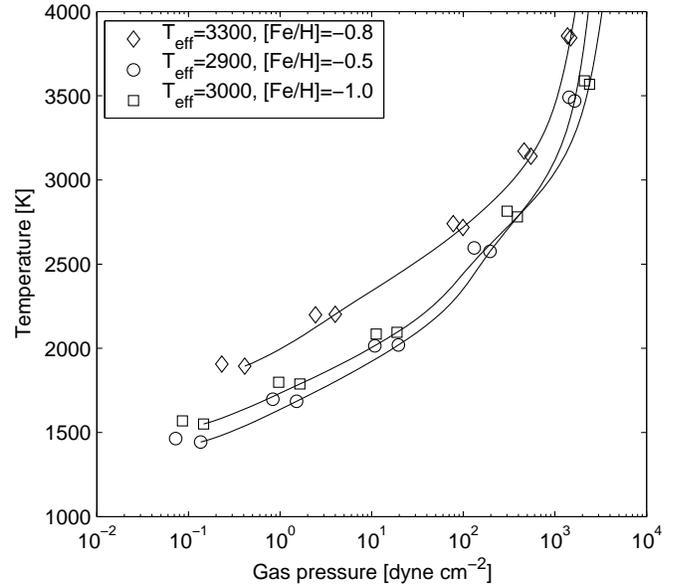}}
  \end{center}
\caption{
Temperature structures of the model atmospheres computed with
the stellar parameters corresponding to the three stars studied
in this work. The points where the optical depth $\log{\tau_\mathrm{Ross}}=-4,-3,\dots,0$ are
indicated by markers. The models corresponding to IGI95-C1 are
marked with diamonds, IGI95-C3 with circles and BMB-B~30 with
squares. The structure is very sensitive to the carbon excess. The
structures computed with the lower value of C/O in Table~3 are
plotted with solid lines and markers while the structures with high
C/O are indicated with markers only.}
\label{fig:atm}
\end{figure}

A  full  discussion  of  the  sources  of  error  in  the  derived  absolute
abundances  and element  ratios due to  uncertainties  in the
atmospheric parameters,  continuum location ($\pm 5\%$)  and random errors  when a
given element  is represented by a few  lines, as in our  case, can be
found  in  Abia \& Isern (1996) and Abia  et  al.  (2001,  2002)  and  will  not  be  repeated
here. These errors  range from $\pm 0.2$~dex for Ba  to $\pm 0.45$~dex
for Ce. We estimate a typical uncertainty in the mean metallicity, [M/H], 
of $\pm 0.30$~dex. The typical error for the elemental ratios 
with respect to the metallicity
([X/M]) range between 0.30-0.40~dex, since some of the uncertainties
cancel out when deriving this ratio. For the same reason, the error
in the abundance ratio between elements ([X/Y]) is somewhat lower, $\pm 0.30$~dex.
For  Nb and Gd, the identified  spectral features are very
weak  and blended,  thus in  some cases  we can  only  establish upper
limits. We estimate an uncertainty in the carbon isotopic ratio of $\pm 9$ (see
Abia \& Isern 1996). These numbers do not include possible systematic errors 
as N-LTE effects or an uncertainty larger than $5\%$ in the continuum location
(see below).

The  abundances  derived   in  IGI95-C3  from  the  $4\,850-4\,950$~{\AA}  region, merit  a 
special note  of caution.  We found
many difficulties to fit this spectral region. In fact, in this region
several choices of T$_{\rm{eff}}$,  C/O ratio and metallicity can lead
to  a similar match  of the  observed spectrum.  Finally we  adopt the
parameters shown  in Table~3  as the best  ones for IGI95-C3  (with C/O
$=1.05$) but,  in any  case, we cannot  obtain as good  a  fit as that
obtained for BMB-B~30 and IGI95-C1 in this spectral region. We believe
that  this may  be  caused  by the  presence  of Merrill-Sanford  bands
(SiC$_2$) in  this star. These molecular bands  are identified between
$4\,100-5\,500$~{\AA} in carbon  stars (e.g. Sarre et al. 2000;  
Yamashita  \&  Utsumi   1968 and McKellar, 1947).  In  fact  we  detect 
extra-absorptions around 4\,867 and 4\,906~{\AA}, not present in the other
stars of the  sample, which coincides with the position  of some 
band heads of this molecule. Merrill-Sanford bands are indeed detected in this
spectral  region  in  some galactic  C-stars with  $T_{\rm{eff}}<3\,100$~K
(Bergeat et al. 2001; Morgan et al.
2004),   and our estimated effective temperature   for   this   star
(T$_{\rm{eff}}=2\,900$~K, see  Table~3)  agrees with  this figure.  This
means that the derived abundances in IGI95-C3 from the region 
$4\,850-4\,950$~{\AA}  have to  be  considered  with caution.   Obviously,
Merrill-Sanford bands  do not affect  the abundances derived  in this
star from the other spectral ranges.

\begin{table*}
\begin{minipage}[t]{\columnwidth}
\caption{Main characteristics of the extragalactic C-stars.
We give the stellar name, the effective temperature, the mean
metallicity   [M/H], the carbon over oxygen abundances ratio (C/O),
the corresponding difference between carbon and oxygen
abundances  ($\epsilon$(C)$-\epsilon$(O)), the carbon isotopic ratio
and the range of estimated bolometric absolute magnitudes.
}             
\label{table:2bis} 
\renewcommand{\footnoterule}{}     
\centering          
\begin{tabular}{c c c c c c c}     
\hline\hline       
                   
Star & T$_{\rm{eff}}$(K) & [M/H] & C/O\footnote{The C/O ratios indicate the range of values estimated from the
analysis of the different spectral ranges (see text).}  & $\epsilon$(C)$-\epsilon$(O)&$^{12}$C/$^{13}$C & M$_{\rm{bol}}$\footnote{The first value indicates the
luminosity obtained from M$_{\rm{K}}$ according to the calibration by Alknis et al. (1998). The second one 
was derived after the bolometric correction by Wood et al.
(1983). See Sect.~2 for more details.}\\ 
\hline                     
   IGI95-C1 & 3\,300 & $-$0.80 & 1.18-1.06&7.23& 25     & $-3.1$, $-3.3$ \\
   IGI95-C3 & 2\,900 & $-$0.50 & 1.10-1.05&7.33&40     & $-4.5$, $-4.5$ \\
   BMB-B~30& 3\,000 & $-$1.00 & 1.20-1.09 &7.35&$>300$ & $-5.6$, $-5.1$\\
\hline                  
\end{tabular}
\end{minipage}
\end{table*}


%
\begin{figure*}
\includegraphics[angle=-00, width=17cm]{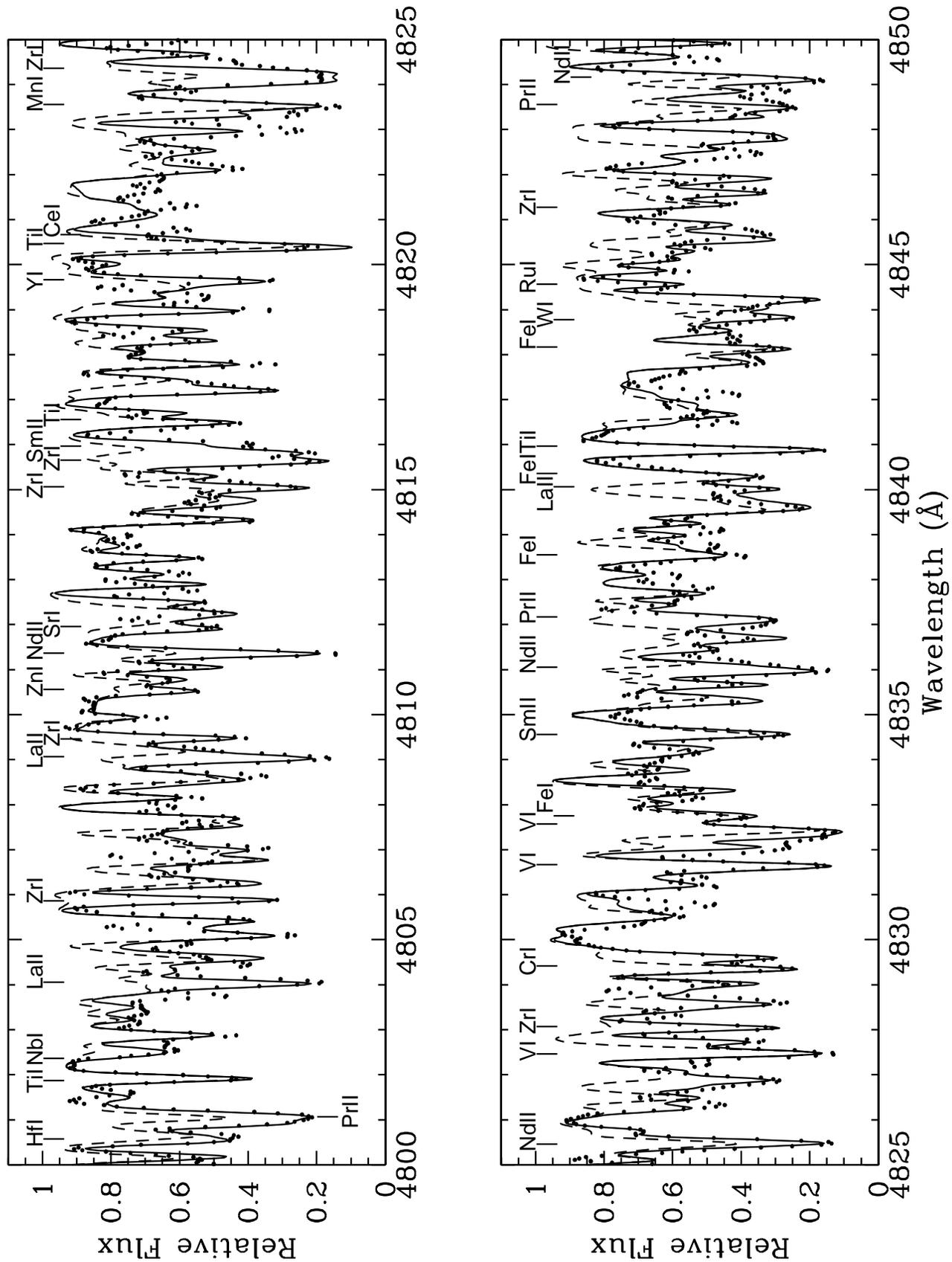}
\caption{Synthetic fits to the spectrum of the star IGI95-C1 in the region around $4\,825$~{\AA}. 
From this region we derived most of the s-elements abundances and the mean metallicity of the stars.
Some atomic lines, main contributors to the specific spectral feature, are marked. Black dots represent 
the observed spectrum. Lines are synthetic spectra: only molecules and metals 
(dashed line), and including s-elements (best fit, continuous line).} 
\label{fig:fit1}
\end{figure*}

\begin{figure*}
\centering
\includegraphics[angle=-90, width=17cm]{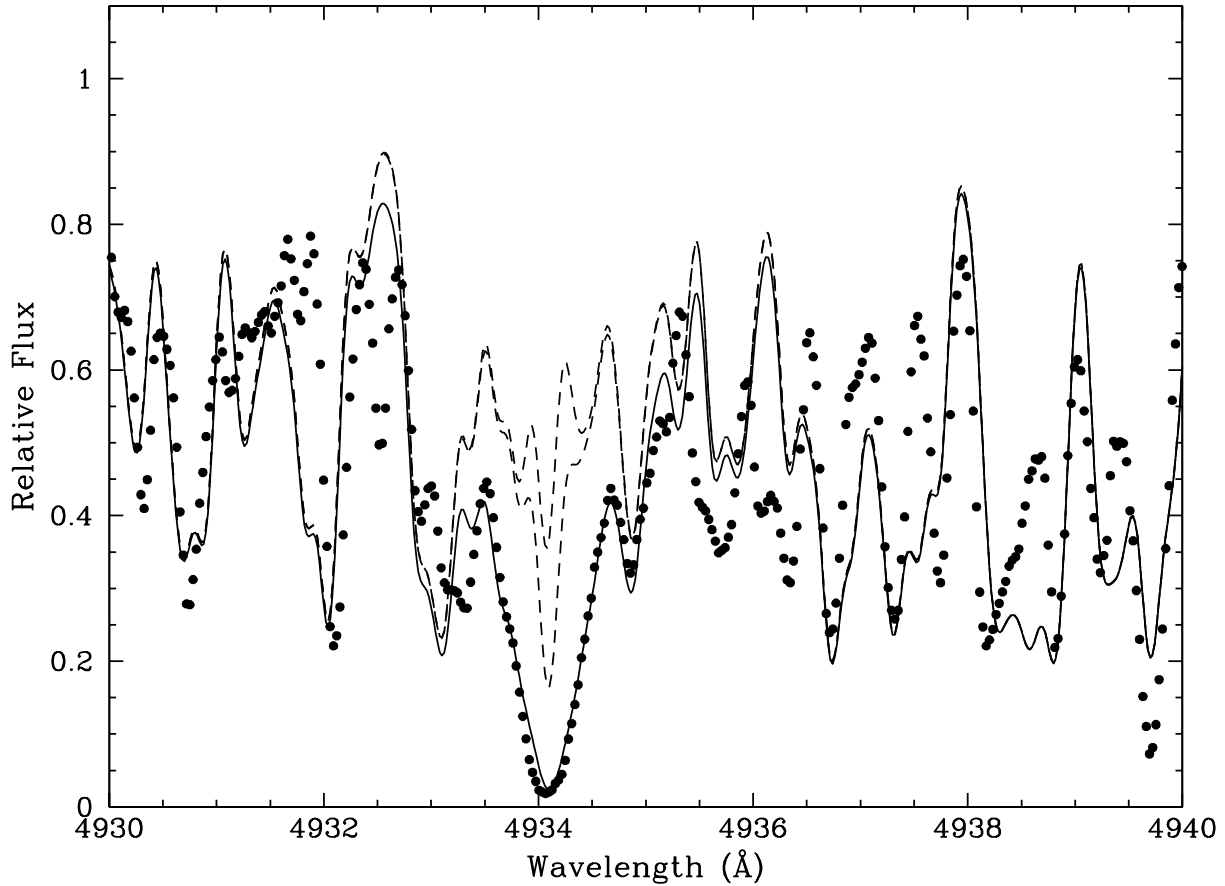}
\caption{Synthetic fit to the spectrum of IGI95-C1 in the region of the Ba II line 
at $\lambda 4\,934$~{\AA}. Black dots are the observed spectrum. Lines are synthetic spectra 
for different [Ba/M] ratios: no Ba and $+0.0$ (dashed lines), $+1.5$ (best fit, continuous line).} 
\label{fig:fit2}
\end{figure*}
\begin{figure*}
\centering
\includegraphics[angle=-90, width=17cm]{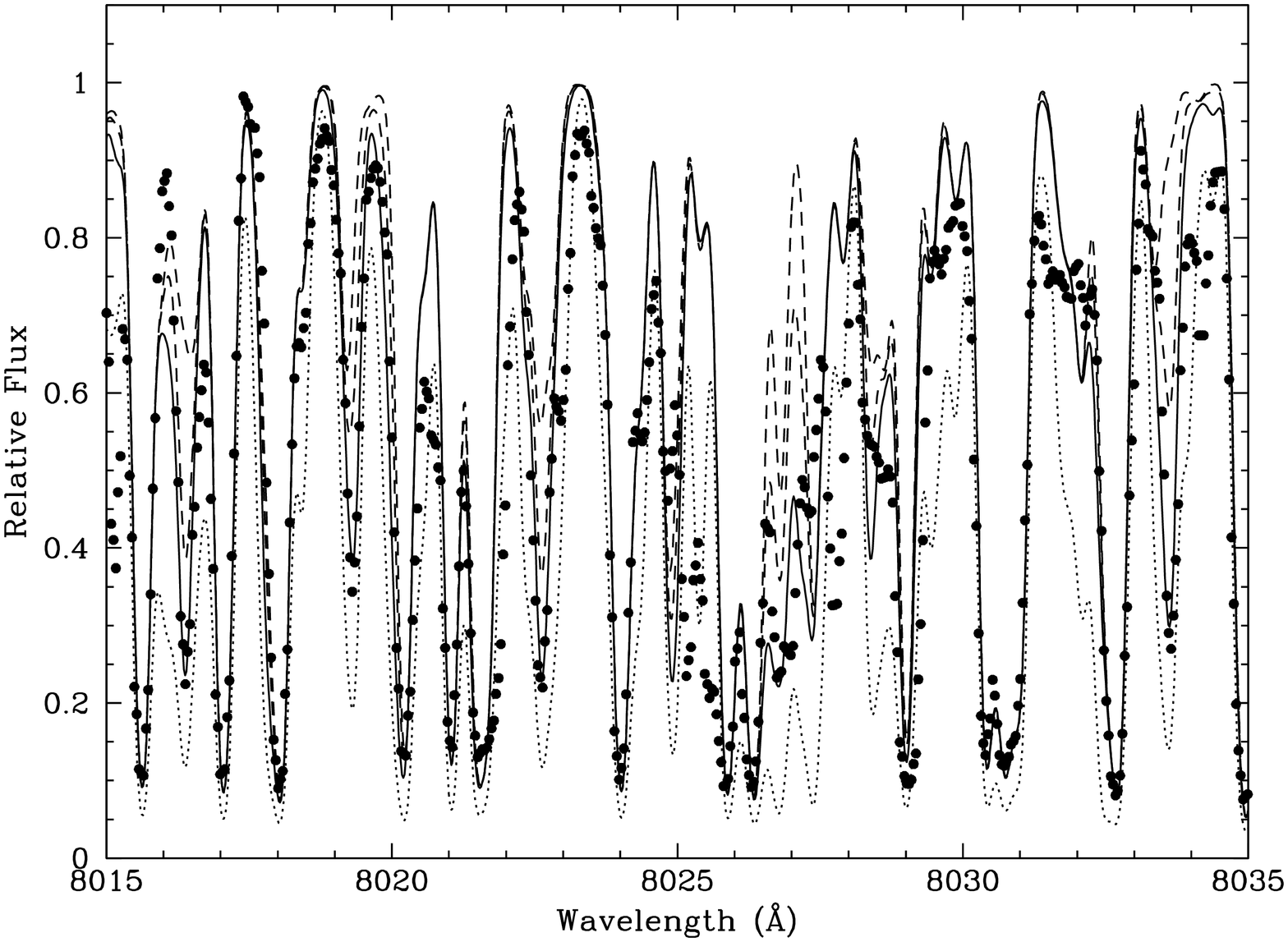}
\caption{Synthetic fit to the spectrum of the star IGI95-C1 in the region around $8\,025$~{\AA} 
from which we derive the $^{12}$C/$^{13}$C ratio and a first estimate of the C/O ratio (see text).
The $^{13}$C abundance is estimated from the weaker $^{13}$CN features at $\lambda\sim 7\,991$, 7\,997, 8\,004, 
8\,016, 8\,019, 8\,023 and 8\,034~{\AA}. The figure shows only some of these features for clarity. As previous figures, black dots 
represent the observed spectrum while lines are synthetic spectra for different C/O and $^{12}$C/$^{13}$C ratios.  
Dashed lines are for no $^{13}$C and  $^{12}$C/$^{13}$C $=100$ with C/O $=1.18$, respectively; 
the continuous line is made with the same C/O ratio and with our best estimate for the $^{12}$C/$^{13}$C ratio in this
star equal to 25.  The dotted line represents $^{12}$C/$^{13}$C $=25$ but with C/O $=1.80$. Clearly, a
C/O ratio much larger than unity is discarded.} 
\label{fig:fit3}
\end{figure*}

\section{Results and discussion}

Fig.~2 and 3 show  examples of theoretical fits to spectral regions
from  where most  s-element abundances  are derived.   The  fits are
reasonably  good  despite the line  lists  used are  likely incomplete. We  
are able to  obtain a reasonable reproduction  of some specific  spectral  
features   representative  of  single  heavy elements. As noted above, for several  
elements the abundance is derived  from just  one line. Table~4 summarizes 
the abundances derived for individual elements in the programme stars. In the case that 
more than one line was used in the analysis we derived the mean value
and its dispersion.

The three stars have low Li abundances, similar to that typically 
found in large Li abundance surveys of galactic C-stars (Denn et al. 1991; Abia
et al. 1993).  The Li content in BMB-B~30 was already studied by Smith
et al. (1995) and their estimate agrees with the value given in Table~4.
On the other hand, the mean metallicity derived in the stars, are compatible with the
typical metallicity ranges derived from the stellar populations of these satellite
galaxies in other observational studies (see references in Sect. 1). In particular, the metallicity 
derived for BMB-B~30, belonging to the SMC, confirms the low values previously reported 
by Rolleston et al. (1999) as the mean metallicity for this galaxy. On the
basis of the derived
Ca and Ti abundances, and considering the error bars, we do not find evidence of $\alpha-$enhancement
in any of the stars studied.

\begin{table*}
\begin{minipage}[t]{\columnwidth}
\caption{Summary of the abundances derived in the programme stars. 
N, indicates the number of lines
utilized for a specific element if more than one line was used (we then
give the mean abundance together with its dispersion). 
[el/M] is the abundance 
ratio with respect to the mean derived [M/H] (see text and Table~3) and is only quoted for 
elements with A$\geq 30$.}
\label{table:3}    
\renewcommand{\footnoterule}{}    
\centering                          
\begin{tabular}{cccccccccc}           
\hline\hline                      
       &   & IGI95-C1      &              &   & IGI95-C3      &                     &  
  &BMB-B~30       &                       \\     
\hline           
Element& N & log $\epsilon$& $\rm{[el/M]}$& N & log $\epsilon$& $\rm{[el/M]}$& 
N & log $\epsilon$& $\rm{[el/M]}$ \\
\hline
Li&  & $<-0.50$     &       &   &$-1.00$      &      &   &$-0.80$      &           \\
Ca&  & --           &       &   &$5.85$       &      &   &$5.30$       &    \\
Ti& 3& $4.20\pm0.12$&       &  2&$4.70\pm0.05$&      &  4&$3.90\pm0.08$&           \\
V&  3& $3.10\pm0.15$&       &  3&$3.50\pm0.15$&      &   & --          &           \\
Cr& 2& $4.90\pm0.10$&       &   &$5.32$       &      &  2&$4.98\pm0.05$&       \\
Mn&  &  --          &       &   &  --         &      &  3&$4.39\pm0.05$&       \\
Fe&  & 6.85         &       &   &  --         &      &  3&$6.50\pm0.08$&       \\
Ni& 2& $5.50\pm0.10$&       &  2&$5.85\pm0.05$&      &   &$5.25$       &        \\
Zn&  & $4.00$       &$0.20$ &   &$4.10$       &$0.00$&   &$4.00$       &$0.40$ \\
Rb&  &$<2.10$       &$<0.50$&   &--           &      &   &$1.40$       &$0.00$ \\
Sr& 2&$3.00\pm0.17$ &$ 0.80$&  2&$3.35\pm0.05$&$0.90$&  2&$2.55\pm0.15$&$0.60$ \\
Y & 2&$2.05\pm0.03$ &$0.60$ &   &$2.55$       &$0.80$&  2&$1.65\pm0.02$&$0.40$ \\
Zr& 4&$2.30\pm0.09$ &$0.50$ &  5&$3.30\pm0.20$&$1.20$&  4&$2.00\pm0.04$&$0.40$ \\
Nb&  &$<0.72$       &$<0.10$&   &$<1.20$      &$<0.30$&  &$<0.42$      &$<0.00$\\
Ru& 2&$1.50\pm0.10$ &$0.50$ &   & --          &       &  &$1.45$       &$<0.60$\\
Ba&  &$2.80$          &$1.50$ &   & $2.55$\footnote{The Ba abundance derived in this star may be affected by 
the presence of Merrill-Sanford bands (see text).} &$0.80$ &  & $2.60$& $1.50$ \\
La& 5&$1.95\pm0.20$ &$1.60$ &   &$1.35$       &$0.70$ & 2&$1.50\pm0.15$&$1.40$ \\
Ce&  &$<2.50$       &$<1.70$&   & --          &       & 2&$1.60\pm0.20$&$1.00$ \\
Pr& 4&$1.10\pm0.04$ &$1.20$ &   & --          &       & 2&$0.80\pm0.08$&$1.10$ \\
Nd& 5&$2.50\pm0.12$ &$1.80$ &  2&$2.00\pm0.05$&$1.00$ & 4&$2.10\pm0.11$&$1.60$ \\
Sm& 2&$1.40\pm0.01$ &$1.20$ &  2&$1.50\pm0.10$&$0.90$ & 2&$1.05\pm0.05$&$0.95$ \\
Gd&  &$<1.50$       &$<1.20$&   &  --         &       &  &  --         &          \\
Hf&  &$1.00$          &$0.90$ &   &  --       &       &  & $1.10$      & $1.20$ \\
W &  &$1.20$          &$0.90$ &   &  --       &       &  & $0.50$      & $0.40$ \\
\hline                                   
\end{tabular}
\end{minipage}
\end{table*}

It should be  remarked that this is  the first time that species like Zn, Ru, Hf 
and W are detected and measured in giant carbon stars. We find a moderate overabundance 
of Zn in all the studied stars. Zn abundance is derived from  the only accessible line 
(Zn I at 4\,810~{\AA}) which seems to be  more sensitive to errors in  the atmospheric 
parameters and damping  constant than other Zn  lines (see  discussion in Chen
et al. 2004).  
In any case, we  estimate a formal error in [Zn/M] of  $0.25$~dex. Keeping in mind this
uncertainty, let us recall that  AGB stars do not produce a sizable
amount of Zn (Bisterzo et al. 2004; Travaglio et al. 2004), so we  can directly compare 
the abundances measured in our sample of  extragalactic carbon stars with that in unevolved galactic
stars.  For [Fe/H] between $-1$ and $-0.5$, unevolved galactic stars show,
on the average, [Zn/Fe] $\sim +0.15$, with a rather large spread 
(Mishenina et  al. 2002; Bihain et al. 2004;  Cayrel et al. 2004; Travaglio et
al. 2004).  Taking into account  this spread and the quoted error bar,
we can  conclude that the overabundance  of Zn we  find in Sagittarius
dSph and SMC C-stars does not significantly depart from the figure found
for galactic stars of similar metallicity. Note
that Reyniers et al. (2004) found [Zn/Fe]$=+0.33$ in the galactic post-AGB star
IRAS 08143-4406 with [Fe/H]$=-0.40$. 

The average abundances  of  the  light s-elements  ($ls$:   Sr,  Y   and  Zr, corresponding to 
the first peak of the main component) and that of the heavier  s-elements ($hs$:  Ba, La,  Nd and  Sm, 
corresponding  to the second peak) are reported  in Table~5. Ce is not included in 
the $hs$ definition,  because there is no single feature in the spectral ranges studied whose 
main contributor is  Ce and, our derivation of its abundance is quite uncertain. 
In addition, we do not include Sm in the case of IGI95-C3 because it was not measured.
The abundance ratios in Table~5 would be not significantly modified if a
different continuum location would have been adopted. 
For instance, we checked 
if a veil of moderate intensity of some unknown molecular 
absorption is contributing in this region. 
It has been found that, considering a
continuum placement $\sim 15\%$ higher, the mean metallicity
should be increased by $+0.25$ dex, the [s/M] ratio by $+0.30$ dex, $+0.25$ dex
for the [ls/M] ratio, $+0.38$ dex for the [hs/M] ratio, but only by $+0.13$ dex
in the [hs/ls] ratio. In this case, however, the global fit obtained to the
observed spectra in the 4\,750-4\,950~{\AA} region is considerably worse. 
The same test shows that the C/O and $^{12}$C/$^{13}$C ratios would be altered
by $+0.02$ dex and $-5$, respectively.

\begin{table}
\begin{minipage}[t]{\columnwidth}
\caption{Metallicity and the s-process indices of the extragalactic C-stars
  studied. The typical error in $\rm{[M/H]}$  and $\rm{[hs/ls]}$ is 
$\pm$0.3dex and the average error on the other indices is  $\pm$0.35dex.
}             
\label{table:4} 
\renewcommand{\footnoterule}{}     
\centering          
\begin{tabular}{c c c c c c }     
\hline\hline       
                   
Star            & $\rm{[M/H]}$  & $\rm{[s/M]}$ & $\rm{[ls/M]}$ & $\rm{[hs/M]}$ & $\rm{[hs/ls]}$\\ 
\hline                     
IGI95-C1 & $-$0.8 &1.1   & 0.6   & 1.5   & 0.9  \\
IGI95-C3 & $-$0.5 &1.0   & 1.0   & 1.0   & 0.0  \\
BMB-B~30 & $-$1.0 &0.9   & 0.5   & 1.3   & 0.8  \\
\hline                  
\end{tabular}
\end{minipage}
\end{table}

Table~5 shows that the three stars have moderate s-element enhancements, [s/M]$\approx +1.0$.
This level of enhancement agrees with the figure found in galactic extrinsic 
(MS, S and CH types with no Tc) and intrinsic (halo) AGB stars of similar metallicity 
(see Busso et al. 2001 and references therein). The same happens when comparing the intrinsic index [hs/ls],
used to characterize the neutron capture process. Hence, independently of their extrinsic or 
intrinsic nature (see below), the extragalactic C-stars studied here are similar to 
their galactic counterparts as far as the s-element enhancements 
are concerned. On the other hand, comparisons with galactic 
post-AGB stars of low metallicity (Zacs et al. 1995; Reddy et al. 1999; Van Winckel \& Reyniers 2000; Reyniers et al. 
2004) show a s-element enhancement typically lower by one order of magnitude. 
  
\subsection{C and O abundances}

Fig.~4 shows a theoretical fit of the observed spectrum for the star
IGI95-C1 in  the spectral  region around  8\,025~{\AA}. This
spectral region is dominated by CN absorptions and it has been used to
derive the C/O  and (mainly) the $^{12}$C/$^{13}$C ratios. The values
obtained for our stars are reported in Table~3. Similarly to the figure  found in most galactic
N-type C-stars (Lambert  et al. 1986; Eglitis \& Eglite  1995, 1997; Abia et
al.  2002), the  derived  C/O  ratios are  only  slightly larger  than
1. Even   considering  the   uncertainty  in   the   model  atmosphere
parameters\footnote{As  commented  in the  previous  section, the derived  C/O
ratios vary depending  on the spectral region analyzed, which
probably indicates  that something is  wrong in the  model atmospheres
and/or molecular line  lists. Note also the possibility of a systematic error
due to the uncertain continuum location.}, we do not believe  that the spectra of
our  stars could  be reasonably  fitted with  a C/O $\sim2$,  or larger as it is
clearly seen in Fig.~4. 
At  solar metallicity,  several  TP and  TDU
episodes  are required to  transform an oxygen-rich M giant  
into a  carbon star, because of the relatively  large O  content in  the envelope.   When the
metallicity is  reduced down  to 1/10 of  the solar value,  C/O $>1$ is
attained after 1 or 2 dredge-up episodes, only. More dredge-up episodes
may be required if the original composition  was enhanced in
oxygen, a possibility that we cannot exclude given the low metallicity
of the stars\footnote{The opposite effect would result if the initial C, N and 
O abundances in the  theoretical models are scaled according to the revised 
solar values by Asplund et al. (2005).}.

It is  often claimed that  the reason for  the recurrence of
C/O $\sim 1$  in the majority of galactic giant carbon stars may be
an observational  bias: an  excess of  carbon in  the envelope
is immediately translated  into a copious production  of carbon-rich dust
that induces  the formation of  a thick circumstellar  shell obscuring
the photosphere  at optical wavelengths.  In this case, the  mass loss
rate can reach up to $10^{-5}$ M$_\odot$/yr and the AGB  star rapidly
looses its  H-rich envelope.  Then, {\it visible} C-stars represent a  
short stage of  AGB evolution.  
We also note that even without the star being obscured by dust, the rapid 
increase in visible and IR gas opacity when C/O passes unity 
might in itself create a
strong levitation which will facilitate the (wind driven) mass loss
and quickly bring the stellar life time to an end.
Numerical calculations (J{\o}rgensen \& Johnson, 1992)
give some support to this theory.
However, exceptions seem to exist: Draco~461 is an intrinsic metal-poor C-star 
belonging to the Draco Spheroidal galaxy which shows C/O $=3-5$  
(Dom\'\i nguez  et al.  2004). Therefore, if we apply the same reasoning to our sample stars,
we should conclude that they have suffered a limited number of
TDUs. 
This might be in contrast with the  evidence that also for
metal-poor  stars,  numerous   TDU  episodes  are  required  to  obtain
s-element enhancements at the level found in the stars analyzed here
(see Table~5). 
In any  case, this  figure merits further 
observational studies in  a larger sample of metal-poor C-stars\footnote{On the other hand, 
it would be interesting  to derive the C/O ratios in a large
sample of CH stars; the only metal-poor (extrinsic) carbon stars where
the C/O ratios found are considerably larger than unity (Vanture 1992).}.

Another possible explanation  of the low C/O may be  a partial conversion of
the  C dredged  up into  $^{13}$C and  $^{14}$N at  the bottom  of the
convective envelope. Intermediate mass AGBs are known to experience a hot bottom
burning (e.g. Lattanzio \& Forestini 1999), but  in low mass AGBs the  
temperature at the base of the convective  envelope is too low to activate  
the CN cycle. Some low  mass  AGB, however,  show  evidences  of  a cool  bottom  process
(Abia et al. 2002; Nollet et al. 2003), i.e. a slow mixing process acting  
below the convective envelope that may be capable of  reducing the C/O  and  the
$^{12}$C/$^{13}$C ratios. An alternative  solution for the low  C/O ratio would arise 
in  the case that our stars  were  extrinsic C-stars. Abia  et al.  (2002) showed
that, under different assumptions for the dilution factor\footnote{The  dilution factor  
is defined  as the  ratio of  the mass transferred  from  the  primary  star  to the  
envelope  mass  of  the accreting secondary star, supposed to  have the original or post first
dredge-up  composition.} the secondary  star cannot appear  as a carbon
star (C/O $>$1)  if the metallicity  exceeds [Fe/H] $\sim-0.3$.  This
also implies that extrinsic  C-stars with a metallicity slightly lower
than this limit  should have C/O ratios not too much  larger than unity, in
agreement  with our finding.   This is  not a  proof of  the extrinsic
nature of our sample stars,  but simply indicates that this hypothesis
is compatible with the measured C/O ratios. We will come back to this point later,  but
note that according to the estimations by Abia et al. (2002) (see their Table~6), 
this hypothesis requires that the C/O ratio in the primary star at the mass-transfer
epoch should have been larger than about 5 (the exact value depends on the metallicity).
As far as we know, only a few planetary nebulae has already been
observed with such a large C/O ratio. 

A second  remarkable result,  which  is related  to the  C/O ratio,
concerns the $^{12}$C/$^{13}$C ratio. In  two of the stars studied here
we find relatively low carbon isotopic ratios ($\approx 20-40$), similar
to  those  found  in  many  galactic  carbon  stars  of  nearly  solar
metallicity (see e.g.  Abia et al.  2003, and  references  
therein).  The observed $^{12}$C/$^{13}$C ratio in galactic RGB stars of low metallicity
is near 7 (Gilroy \& Brown 1991; Gratton  et  al. 2000).  As  a consequence  of  the  TDU, the  carbon
isotopic ratio is expected to increase during the TP-AGB phase.  Then,
the values  measured in the case  of IGI95-C1 and IGI95-C3,  namely about 5
times larger than  the typical value found in  bright red giant stars,
are  compatible  with  the   $^{12}$C  enhancement  implied  by  their
C/O $\approx$1.  This  is also  valid  in  the case  of mass  transfer  and
dilution (extrinsic C-stars). 
In contrast to this, the very high carbon isotopic ratio found in 
BMB-B~30 ($>300$) 
would require a much higher C/O ratio than the one we derive.
There are no obvious solutions for
this problem,  namely how to keep  the C/O ratio close  to unity without
decreasing the $^{12}$C/$^{13}$C ratio.  A possibility would be
the existence of a deep enough convective intershell region during the thermal
pulse. This would mix O (in addition to C) up to the top of the intershell.
Then, the subsequent {\it normal} TDU may bring C and O to the surface, reducing
the increase of the C/O ratio. However, there are at least two important consequences
of this. First, because the temperature at the base of the convective TP would be
as large as $3.5\times 10^8$ K, then one would expect an s-element pattern altered
according to the activation of the $^{22}$Ne neutron source. Second, since the TDU of
carbon would be very strong, the $^{12}$C/$^{13}$C ratio should be very large ($\sim 500-1\,000$).
Leaving apart that this possibility may have also important consequences for the subsequent 
evolution of the star, the s-element pattern found in this star (see Table~4,
5
and Fig.~5) 
is at odds with this possibility.

From  the previous discussion, it is seen that  the C-N-O  abundances are
key to understanding AGB evolution and nucleosynthesis.  Additional
observational  studies   in a large  sample  of   stars  are  required,
preferably with  different indicators, such as the CNO-bearing molecular
bands in the  infrared, which seem to  be less prone to
line formation problems like saturation or blends.

\begin{figure*}
\centering
\includegraphics[angle=-90, width=15cm]{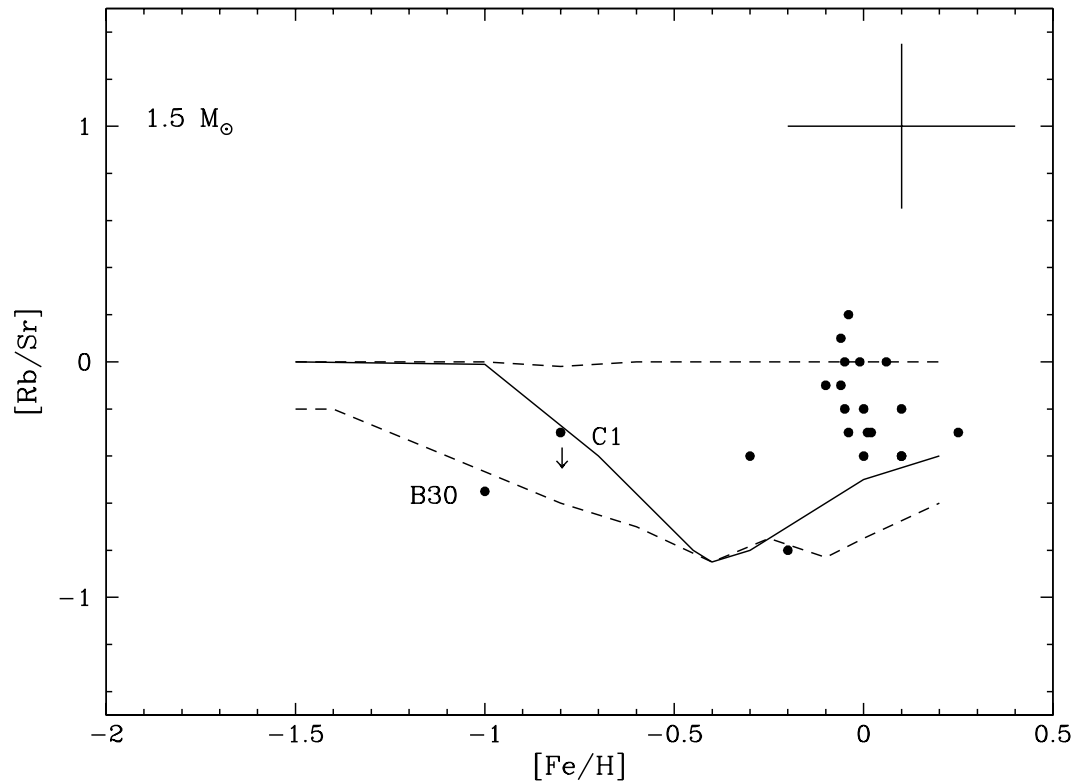}
\caption{Comparison of the observed [Rb/Sr] ratio vs. [M/H] with the 
theoretical predictions for different choices of the $^{13}$C-pocket (see text) 
for a 1.5 M$_\odot$ TP-AGB. Continuous line represents the ST (standard) choice while 
dashed lines limit the area allowed by the choices from STx2 to ST/12. Theoretical
models are from Gallino et al. (1998). The stars studied here 
are marked. The abundance ratios corresponding to IGI95-C3 are not shown because we do not 
detected Rb in this star. Similar ratios derived in galactic N-type carbon stars (Abia et al. 
2001) (black dots at near solar metallicity) are also shown for
comparison. } 
\label{fig:rbsr}
\end{figure*}
\begin{figure*}
\centering
\includegraphics[angle=-00, width=15cm]{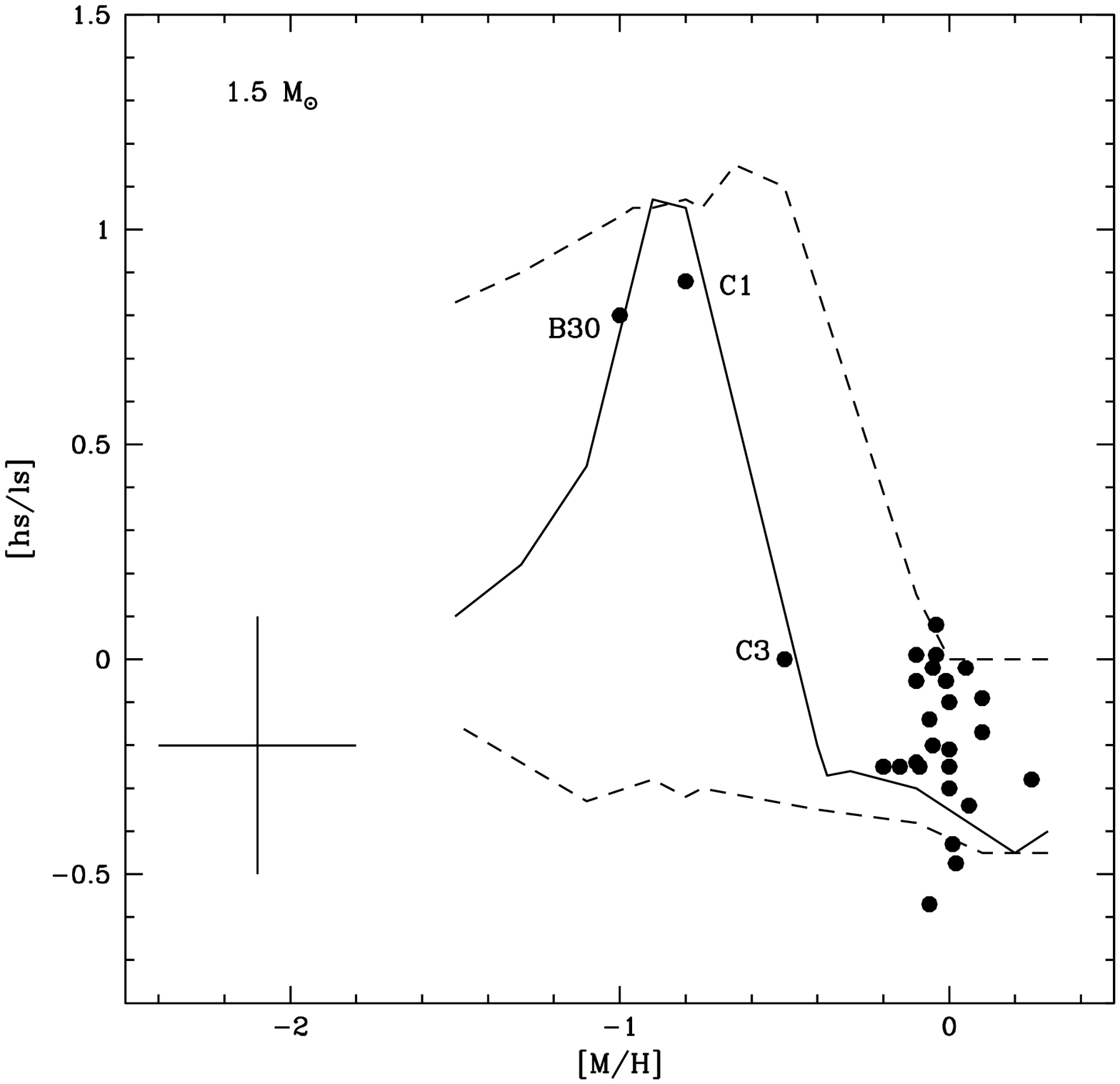}
\caption{Comparison  of   the  observed  mean of heavy (Ba, La, Nd and Sm) to 
light-mass (Sr, Y and Zr) s-element [hs/ls] enhancement  (signature of the 
neutron exposure) against metallicity  with theoretical prediction for 
a 1.5~M$_\odot$ TP-AGB. Continuous and dashed lines represent the same as 
Fig.~\ref{fig:rbsr}. 
The location of the stars studied here are marked. At near solar metallicity, the galactic carbon stars studied in 
Abia et al. (2002) (after minor revisions) are shown for comparison.}
\label{fig:hsls}
\end{figure*}

\subsection{Heavy elements}
  
The  relative abundances  of  s-elements that  follow the  $^{85}$Kr
reaction  branching (Rb,  Sr, Zr  and  Y) along  the s-process  path
yield  information  on  the  neutron density  prevailing  during  the
neutron  capture processes  occurring in  the He  intershell  (Beer \&
Macklin 1989; Lambert et al.  1995; Abia et al. 2001).  In particular,
the   [Rb/Sr]    helps to   discriminate
between   the  two   main   neutron  sources,   the
$^{13}$C$(\alpha,n)^{16}$O and $^{22}$Ne$(\alpha,n)^{25}$Mg, operating
in TP-AGB  stars. The former  
dominates  the s-process  nucleosynthesis  in low  mass AGBs  
(M~$<3$M$_\odot$) for which [Rb/Sr] $<0$ is expected.  
Fig.~5 shows the  comparison of the [Rb/Sr] ratios derived 
in BMB-B~30 and IGI95-C1 (no detection for IGI95-C3)
with s-process model predictions for a  
1.5~M$_\odot$ AGB star at C/O~$>1$ (Gallino  et al. 1998; Busso
et  al. 1999; Busso  et  al. 2004).  The  same  ratios derived  in
galactic N-type  C-stars  are also shown  for comparison.  The 
lines  in  Fig.~5
correspond  to  a specific assumption about  the
$^{13}$C  abundance  in the  $^{13}$C-pocket.   We  show  the  case
defined as  standard (ST) by  Gallino et al. (1998),  corresponding to
$4\times 10^{-6}$ M$_\odot$ of $^{13}$C\footnote{For a more 
detailed discussion  
of the models see Gallino et al. (1998).}. The derived ratios are in good
agreement with the model predictions and are similar to those obtained 
for galactic N-type C-stars. Comparison of the derived [Rb/Zr,Y] ratios with
model predictions leads to the same conclusion.

Fig.~6 reports the intrinsic index [hs/ls] ratio in our stars 
in comparison with model predictions for a 1.5  M$_\odot$ AGB star at the
beginning of the TP-AGB phase (the same models as Fig.~5).  
The solid line 
represents the standard choice for the amount of $^{13}$C in the $^{13}$C-pocket (ST case). Similar 
abundance ratios found in N-type galactic C-stars (Abia et al. 2002) are 
also plotted.  At high (nearly solar) metallicity the
light-s (ls)  are slightly  overproduced with  respect to  the Ba
group elements (hs). The [hs/ls] ratio increases with decreasing metallicity
as more neutrons are available  per seed. A maximum is reached around
[Fe/H] $\sim -1$, a metallicity within the range spanned by our sample of
stars.  Then, at  very low  metallicity ([Fe/H] $\leq -2$) and at the end of the
TP-AGB phase,  the [hs/ls] tends toward  a constant value,  since most of the  neutrons 
are spent  to produce lead (see e.g. Busso et al.  1999). 
The three studied stars seem to follow
the  variation with  the metallicity  of  the standard  choice of  the
$^{13}$C-pocket (ST case).  Obviously, the large uncertainty in the 
abundance ratios and the scarcity of data at low metallicity prevent us 
from concluding whether 
there is  a preferred {\it average} abundance of $^{13}$C (or spread) 
with a  different effect at different  metallicities. 
   
To return to the  question of the extrinsic or intrinsic nature
of  our stars, as  it is well  known,  the detection  of  Tc is  an
undeniable evidence that a  star, with s-element enhanced composition,
is currently undergoing TDU. The only Tc isotope with a
sufficiently long life to be  observable is indeed $^{99}$Tc, whose
decay  lifetime\footnote{The  value  is  $\tau_{1/2}\sim
2\times  10^5$ yr.}  is  comparable  to the  time  elapsed between  two
subsequent TPs in low mass  AGB stars. This Tc isotope is produced by
the s-process and  it  is expected  to  appear at  the  surface of  a
TP-AGB   star  undergoing  TDU.   Unfortunately,  our
instrument set-up  did not include the adequate  spectral regions for
 Tc  detection.  However, an alternative test  of the intrinsic  nature of a
C-star is  provided by  the measurement of  Nb abundance, a chemical element with 
only  one stable  isotope, $^{93}$Nb,  which  can be  produced by 
$^{93}$Zr decay\footnote{Half-life $\tau_{1/2}\sim  1.5\times 10^6$
yr.}.  The $^{93}$Zr  lifetime is comparable with the  duration of the
whole AGB phase and, thus, the  comparison of the abundance of Nb and
that of  its neighbors  (e.g. Zr,  Y) would tell  us if  the s-process
enhancement is intrinsic or due to mass transfer.
S-process   nucleosynthesis  calculations
predict that  the Zr isotopes  and, in particular  $^{93}$Zr, increase
during the AGB  phase, while Nb remains almost unchanged (Straniero et al. 2005). As a consequence,
when the  star attains  the C-rich star stage, [Nb/Zr] $\sim  -0.5$, almost
independently of its metallicity.  After a few million years, however, as a consequence
of the $^{93}$Zr decay, this ratio rises to the scaled solar value.
In  other words,  for  an extrinsic  C-star [Nb/Zr]  should be  $\sim
0$.  From  our spectra  we  were only  able  to  derive upper  limits,
[Nb/Zr]$<-0.4 \pm 0.4$, for  all the  three studied C-stars. This  figure would
hence  indicate that  they are  probably intrinsic  C-stars.  A better
determination  of the  Nb  abundance  is however needed to give  a
definitive answer.

\section{Conclusions}
In this paper we have presented for the first time a detailed chemical
analysis of three extragalactic carbon stars, focusing 
mainly on their s-element content and its dependence on metallicity. The three stars are moderately
metal-poor with a mean metallicity compatible with the typical metallicity
observed in the main stellar population of the respective host satellite galaxy, namely: the
SMC and Sgr dSph. All the stars show an s-element enhancement of [s/M]$\approx +1.0$,
which is of the same order as that found in galactic AGB stars of similar metallicity.
The s-element abundance pattern found across the $^{85}$Kr branching point and 
the second s-peak (Ba group elements) 
of the s-path can be understood on the basis of 
theoretical nucleosynthesis models, if the stars have 
low masses (M$< 3.0$ M$_\odot$),
and the $^{13}$C$(\alpha,n)^{16}$O reaction is
the main source of neutrons.
The derived intrinsic index [hs/ls] in the three stars 
seems to be in agreement with  
a preferred choice for the abundance of $^{13}$C that is burnt in stars with 
different metallicity. However, 
given the large error bars in the abundances and
the limited number of stars analysed here, our results are not in 
contradiction with the figure extracted from studies in galactic 
AGB \& post-AGB stars concerning the existence of a dispersion in the 
amount of $^{13}$C burnt in stars with different metallicity. 
Obviously, more extended studies with narrower error bars
on the derived abundances are needed to confirm this.
As for galactic C-stars, the C/O ratios derived in these three 
extragalactic C-stars
are found to be only slightly larger than unity. 
This fact, together with the derived  $^{12}$C/$^{13}$C ratios
and the large s-process enhancement, is in
disagreement with
theoretical predictions for TP-AGB stars.    
   
\begin{acknowledgements}
We warmly thank B. Gustafsson for his careful reading of the manuscript
and his valuable suggestions that considerably improved this work.
The referee, H. Van Winckel, is thanked for his useful comments.
This publication makes use of data products from the Two Micron All Sky
Survey, which is a joint project of the University of Massachusetts and the
Infrared Processing and Analysis Center/California Institute of Technology,
funded by the National Aeronautics and Space Administration and the National
Science Foundation. Part of this work was supported by the Spanish grant 
AYA2002-04094-C03-03 from the MCyT. 
\end{acknowledgements}

\end{document}